\documentclass[preprint2]{aastex}

\def\easca{\emph{ASCA}\ }
\newcommand\farcsec{\mbox{$.\!\!^{\prime\prime}$}}%
\def\simlt{\lower.5ex\hbox{$\; \buildrel < \over \sim \;$}}
\def\simgt{\lower.5ex\hbox{$\; \buildrel > \over \sim \;$}} 

\def\cm3{{\rm cm^{-3}}}
\def\kms{km s$^{-1}$}

\def\nh{N_{\rm H}}
\def\nhone{N_{\rm HI}}
\def\nhtwo{N_{\rm H_2}}

\lefthead{Koo et al.}
\righthead{AN \easca STUDY OF THE W51 COMPLEX}

\begin{document}

\title{ AN \easca STUDY OF THE W51 COMPLEX} 

\author{Bon-Chul Koo and Jae-Joon Lee}
\affil{Astronomy Program, SEES, Seoul National University,
Seoul 151-742, Korea;\\
koo@astrohi.snu.ac.kr}
\author{and \\ Frederick D. Seward}
\affil{Harvard-Smithsonian Center for Astrophysics, 60 Garden Street, Cambridge, MA 02138}

\begin{abstract}

We present the analysis of \easca archival data from the Galactic source W51. 
The \easca spectra show that the soft ($kT\simlt 2.5$ keV) X-rays are of thermal
origin and are compatible with W51C being a single, isothermal ($kT\simeq 0.3$~keV)
supernova remnant at the far-side of the Sagittarius arm.
The \easca\ images reveal hard ($kT\simgt 2.5$ keV) X-ray sources 
which were not seen in previous X-ray observations. 
Some of these sources are 
coincident with massive star-forming regions and the 
spectra are used to derive X-ray parameters. 
By comparing the X-ray absorbing column density 
with atomic hydrogen column density, we 
infer the location of star-forming regions relative to 
molecular clouds. There are unidentified hard X-ray sources 
superposed on the supernova remnant 
and we discuss the possibility of their association.

\end{abstract}

\keywords{ISM: individual (W51) --- supernova remnants --- H~II regions --- stars: formation 
--- X-rays: ISM}

\section{INTRODUCTION}

W51 is an extended ($\sim 1^\circ$) 
radio source located at the tangential point ($l=49^\circ$)
of the Sagittarius arm. 
It is composed of two complex H~II regions, W51A and W51B, and the 
supernova remant (SNR) W51C (e.g., Bieging 1975; Koo 1997). 
W51A forms the northern part of W51, 
and is separated from the other two sources.
It contains two major components, each composed of several 
compact H~II reigons.
W51B is composed of at least six compact H II regions scattered over
an area of $\sim 15'$ size.  W51B sources are associated with 
a stream of atomic and molecular gases, with
line-of-sight velocities
significantly greater than the maximum velocity permitted
by Galactic rotation alone.
The stream is thought to be gas flowing along the
Sagittarius spiral arm in response to 
the perturbation due to the spiral potential (e.g., Burton 1971).
Superimposed on W51B, there is an extended
structure W51C which is a SNR.
W51C appears in radio continuum as an incomplete shell of $\sim 30'$ extent 
with its upper portion open \citep{cop91, sub95}. Shocked atomic
and molecular gases have been detected in the western part of the SNR, which
indicates that the SNR is interacting with a large molecular cloud
\citep{koo91, koo97a, koo97b}. 
The complex structure of W51 is partly due to the inclination of 
the Sagittarius arm, so that we 
look down the length of the arm a distance of 
5~kpc along the line-of-sight.

Soft ($\simlt 2$ keV) 
X-ray emission associated with W51 has been detected by {\em Eintesin}
and ROSAT 
\citep[hereafter KKS]{sew90, koo95}.
Diffuse X-rays come from a region surrounding W51B and W51C, whereas
W51A has essentially no soft X-ray emission associated with it.
The soft X-ray emitting region is elongated 
($50'\times 38'$) along the east-west direction, 
and may be divided into three parts;
a central structure composed of two bright regions separated by $\sim 10'$,
an incomplete shell $\sim 30'$ long in the east,
and an extended ($\sim 20'\times 10'$) structure in the west.
The structure in the west is separated from the central region 
by a region of weak emission where 
molecular clouds associated with the W51B star-forming region   
are located.
The systematic hardening of the X-ray spectrum toward the west 
suggested that the X-rays are emitted behind these molecular clouds.
The average X-ray spectrum of the SNR was fitted well by
a single-temperature ($k_B T\simeq 0.29$~keV) thermal plasma model.  
But the energy range of the ROSAT detector was not enough to 
distinguish this from other emission models.

In this paper, we present the results of an \easca study of the W51 complex. 
The greater (0.7--10 keV) energy coverage of \easca makes it clear that 
the X-ray emission from the SNR is of thermal origin. The newly derived 
plasma parameters are consistent with the ROSAT results. 
\easca also reveals hard X-rays from star-forming regions in W51, which were
not seen by ROSAT.

\section{\easca Data and Image Processing}

W51 was observed by \easca October 10, 1995. Two chips 
of each SIS detector were aligned in 2-CCD mode 
to cover the bright central region revealed by ROSAT, while the GIS 
detector covered 
most of the X-ray emitting region. Figure~1 shows the positions of 
SIS and GIS fields overlaid on the ROSAT image of KKS.
The data used are the archive data processed by 
the revision 2 standard screening process.
The total exposure times after screening were 36.4~ks and 12.8~ks 
for SIS0 and SIS1, and 44.3~ks for both GIS2 and GIS3. 

For spatial analysis, we have generated GIS images in the energy 
bands: 0.7--1.5 keV, 1.5--2.5 keV, 2.5--6.0 keV, and 6.0--10.0 keV. 
We first generated a raw GIS image for a given energy band and 
divided it by the blank-sky GIS image of the same energy band.  
The blank-sky image represents the response of X-ray telescope (XRT) 
plus GIS for uniform background emission. 
Therefore, by dividing the raw image by the blank-sky image, 
the background level in the image could be flattened (e.g., Kaneda 1998),
and by subtracting a constant, a background-subtracted image was obtained. 
For a test, we applied the technique to 
a source-free region at $\ell\simeq 50^\circ$ and 
confirmed that the residual image did not have any systematic gradient. 
The vignetting effect for spatially-confined sources, however, is greater 
than that for the background because there is no 
stray-light coming from outside the field-of-view.   
We therefore divided the above image 
by an effective exposure map which accounts for the 
variation of effective area with detector position and energy. 
We constructed GIS2 and GIS3 
images separately and 
added the two after masking out regions near the edge of the field 
and around the built-in calibration source. 
The images have 
pixel size of $15''\times 15''$ and have been 
smoothed using a Gaussian profile of $\sigma=1'$. 
The count rates  
in the GIS2+3 images of 0.7--1.5, 1.5--2.5, 2.5--6.0, and 6.0--10.0 keV 
are 0.51, 0.64, 0.46, and 0.26 counts s$^{-1}$, respectively. 
Images of wider energy bands, e.g., 0.7--2.5 keV, were constructed 
by adding the above images. 
We generated SIS images in a similar way, 
except that they were not divided by blank-sky images 
because the background intensity was uniform over the field.
SIS images have pixel size of $6\farcsec 4 \times 6\farcsec 4$ and have been 
smoothed using a Gaussian profile of $\sigma=25\arcsec$. 

\section{Spatial Analysis}
\subsection{\easca Images}
\subsubsection{GIS Images}

Figure~2a is GIS image of W51 in the total (0.7--10.0 keV) energy band 
and shows that the region has a complex X-ray morphology. 
This is partly due to the superposition of 
unrelated sources, which can be distinguished by looking at 
soft and hard X-ray images.
Figures~2b and 2c are GIS images of W51 in the soft (0.7--2.5 keV) and hard 
(2.5--6.0 keV) energy bands.
Figure 2b, which is consistent with the ROSAT image of 
KKS (Fig. 1), shows an extended structure
$\sim 20'$ long along the SE-NW direction at the field center. 
It is composed of two bright 
regions, regions XN and XS, separated by $\sim 10'$. XS has a ridge 
extending to the east. There is also a bright 
extended structure in the west, XW. 
These structures are immersed in fainter diffuse emission.
 
In hard X-rays 
(Fig. 2c), there are also bright regions and faint diffuse emission. 
The bright regions are small and could be unresolved sources. Some of the diffuse 
emission might be instrumental because 
the point spread function of the \easca XRT has a broad wing.  
Excluding emission at the edge of the field, 
we identify three sources. 
Two of them (HX1 and HX3) are somewhat extended and might have 
more than one component; in HX1, there appears to be a 
structure to the southeast of the maximum, while HX3 appears double. 
Sources HX2 and HX3 are located close to regions XS and XN, respectively, and 
the deformation of the contours 
in Figure 2b corresponds to these hard sources. 
Table 1 summarizes source properties. HX3-east and HX3-west represent 
the east and west components of HX3, respectively. 
The peak positions and count rates 
are derived from Figure 2c. 
The positional uncertainty is $\sim 1'$. 
The count rates are uncertain 
($\simgt 20\%$) because of background subtraction, confusion with 
other sources, and the vignetting correction.  
The last four columns 
summarize the properties of the radio counterparts (see \S~3.2 ).

Figure 2d is an X-ray color map generated from three energy bands: 
0.7--1.5 keV (red),
 1.5--2.5 keV (green), and 2.5--6.0 keV (blue), and offers an alternate 
way to look at this region. The 
soft sources in Figure~2b appear with
different colors, regions XN and XS in yellow, region XW in green, 
and the faint features in the southeast in red. This illustrates the 
result of KKS who found that the X-ray spectrum between 0.6 keV and 
2.2 keV becomes systematically harder toward the west.  
As will be shown in \S~3.3, this is probably due to the absorption 
of intervening interstellar gas.
The hard X-ray sources of Figure~2c appear blue in Figure~2d.

\subsubsection{SIS Images}

Figures 3a and 3b are the soft and hard SIS images of the central region of 
W51 (cf. Fig. 1). 
These have higher ($\sim 1'$) resolution and show more structure. In order to 
help compare Figures 2 and 3, we mark the peak positions in Figures 2b and 2c as 
crosses in Figures 3a and 3b, respectively.
According to Figure 3a, the brightest part of 
region XN is elongated ($\sim 3'$) eastwest, while
the brightest part of region XS
is resolved into northern and southern components. 
In Figure 3b, HX2 is visible in the southern area while HX3-east is barely visible
near the northwestern boundary.
HX2 and the southern component of XS are both elongated eastwest and 
partly overlap, which suggests that they might be 
the same source.

\subsection{Comparison with Radio Distribution}

An overlay of X-ray and radio images helps to understand this complex region.
Figure~4a compares the ROSAT X-ray image of KKS with 
a radio contour map. This ROSAT image 
contains almost the same information as the \easca soft X-ray map 
but covers a larger area. KKS did a similar comparison, but the radio maps 
that they used either had  
poor angular resolution (FWHM=$4.'2$) or did not show 
large-scale ($\simgt 15'$) structures. 
Figure 4a uses the 330~MHz map of \citet{sub95} which shows the 
detailed ($\sim 1'$) structures over the whole complex. 
In Figure~4a, the clusters of compact sources in the north and near the center 
are known as the W51A and W51B H~II region complexes, respectively.
The W51C SNR appears as an incomplete, very thick radio shell in the southeast. 
The eastern X-ray shell matches almost perfectly with the outer 
boundary as expected for a SNR.
Regions XN and XS are partly surrounded by the radio shell, and 
they appear to be immersed in a diffuse X-ray emission 
together with the eastern X-ray shell. 
This might indicate that they are parts of the SNR too. 
SNRs with centrally-brightened X-ray emission are not unusual (see \S~5.1).
On the other hand, the western half of the SNR is not clearly defined in radio
because of the superimposed source W51B. But 
the western edge of the radio source is also matched in X-ray emission with 
region XW.
This probably shows the western shell of the SNR.   
In \S~4.1, we will show that regions XN, XS, and XW are all located 
at the far-side of the Sagittarius arm and have similar temperatures, which 
suggests that they are all possibly parts of a single SNR.

Figure~4b compares the hard X-ray image with the radio image and 
shows that some of the hard X-ray sources  
are associated with compact radio sources. 
First, HX1 coincides with 
G48.9$-$0.3 and G49.0$-$0.3 which are compact H~II regions.
Second, HX3-east coincides with 
G49.2$-$0.3, also a compact H II region.
Third, the bright X-ray source at the
northern boundary of the \easca field might be associated with W51A
which is mostly outside the field of view. 
The hard X-rays from these sources, therefore, might originate from  
early-type stars and young stellar objects. 
For HX2 and HX3-west, there are 
no obvious corresponding radio sources in Figure~3b, although there seems 
to be a weak enhancement at HX2.
We examined the image of the NRAO VLA Sky Survey (NVSS) 
at 21 cm \citep{con88} 
and found that there are indeed weak radio sources coincident with 
these hard X-ray sources too. The nature of these radio sources is unknown.
The results of the comparison are summarized in Table~1.

\subsection{Comparison with CO Distribution}

KKS compared the distribution of ROSAT X-ray surface brightness with that of CO 
and concluded that the spectral hardness between 0.6--2.2 keV was due to the 
absorption of X-ray photons by intervening interstellar gas. 
There is now more sensitive CO data with higher spatial resolution.
Figure 5a compares the ROSAT X-ray image with 
the CO J=1--0 integrated-intensity map generated from the data 
of \citet{car98}. The CO map is obtained integrating over the velocity range 
of 0 to +75~\kms\ and shows all the molecular gas toward this direction.
The CO map has an angular resolution 
of $\simeq 45''$ with 50$''$ sampling. 
W51A is associated with the bright CO cloud in the northern part of 
the field, while
the W51B radio sources are associated with a filamentary cloud 
in the central field. 
(For a detailed comparison of radio sources and molecular clouds in W51B, 
see Koo [1999].)

Figure~5a strongly suggests that 
the spectral hardening in soft X-rays is due to the
absorption of the intervening interstellar gas. 
First there is generally more molecular gas 
toward region XW than regions XN and XS. Second, 
the central filamentary cloud partly overlaps with 
the western boundary of region XN where 
the spectrum is harder (Fig. 2d).  
If we use 
$N_{\rm H_2}/W_{\rm CO}=2.3\times 10^{20}$ molecules cm$^{-2}$ (K km s$^{-1}$)
where $W_{\rm CO}$ is the integrated CO J=1--0 line intensity in 
K~km~s$^{-1}$ and $N_{\rm H_2}$ (cm$^{-2}$) 
is the molecular hydrogen column density
\citep{str88}, then, 
the average H$_2$ column densities 
toward regions XW, XN, and XS are 
0.54, 0.45, and 0.20 $\times 10^{22}$~cm$^{-2}$, respectively. 
The H~I column density is also 
somewhat greater toward the 
west, e.g., 1.6, 1.5, and 1.4$\times 10^{22}$~cm$^{-2}$ toward 
XW, XN, and XS.
As will be shown in \S~4.1, the spectra of the three regions require 
different absorbing column densities of hydrogen nuclei and 
this result is consistent with the X-rays coming from behind 
most of the atomic and molecular gases. 
The red color of the southeastern X-ray shell 
in Figure~2d is consistent with the negligible CO emission and smaller  
(1.2$\times 10^{22}$~cm$^{-2}$) H~I column density in this direction.

Figure~5a also shows that the central filamentary molecular cloud appears to fill
the gap between the central and the western X-ray emitting regions (see Koo [1999]
for a detailed structure of this cloud.). KKS examined their correlation and
concluded that the column density of the cloud is large enough to produce the
dip, but, because we see soft X-rays at the northwestern boundary of XN toward
which $\nh$ is comparable, the intrinsic brightness in the dip must be fainter
than the central region. According to Figure~5a, there are regions of low soft
X-ray emission in the dip where the CO emission is rather weak, and this might
be another evidence suggesting that the intrinsic soft X-ray brightness in the dip
is faint.

On the other hand, Figure~5b shows that the X-ray sources 
HX1 and HX3-east are associated with molecular clouds, which is 
consistent with these sources being compact H~II regions. 
HX2 and HX3-west are not associated with CO emission, which 
suggests that they might be either background sources or 
sources associated with the SNR. (The contour surrounding HX3-west 
in Figure~5b represents a minimum.)

\section{Spectral Analysis}

As we have seen in \S~3, W51 is a complex region with 
X-rays coming from a SNR, star-forming regions, and perhaps background sources. 
In this section, we first 
inspect the spectra of regions bright in soft X-ray, which are 
considered to be the parts of the SNR W51C. 
We then inspect the spectra of hard X-ray sources,  
some of which are star-forming regions and some are unidentified.

\subsection{W51C SNR}
\subsubsection{GIS spectra}

Figure~6 shows the spectra of regions XN, XS, and XW. Each spectrum is obtained
by adding all the photons within the corresponding area marked in Figure~2b
and by subtracting a background spectrum.
The background spectrum in each region was estimated from the source-free 
region in the northern part of the field.
There are several points to be made from Figure~6.
First, the spectra clearly show line features at 
$\sim 1.3$ keV and $\sim 1.8$~keV,
which indicates that the emission is of thermal origin.
The central energies of the line features are close to those of 
the K lines of He-like Mg and Si ions.
Second, the spectra of regions XN and XS clearly have an excess emission
at $E\simgt 2.5$ keV compared to XW. This
excess emission might be due to the hard X-ray 
sources HX2/HX3 but there is also the  
diffuse emission (Figure~2c).
Third, at $E\simlt 2$ keV, the spectrum becomes harder in 
the order of regions XS, XN, and XW, 
which is consistent with Figure 2d.

We first tried to fit the spectrum with 
a single-temperature, equilibrium thermal emission
model. The fitting was done using
the MEKAL code inside the XSPEC package (version 10).
The data between $E=0.7$ and 6.0 keV
were used. Both the gas temperature $T$ and the absorbing column density of
hydrogen nuclei $\nh$ were varied and the goodness of fit to the data was
calculated. The elemental abundances were fixed to be solar.
The fitting, however, was not satisfactory for all three regions.
We therefore fitted a
two-temperature equilibrium thermal model for regions XN and XS. For region XW,
which has low S/N ratio, we instead limited the energy range to 
$E=0.7$--2.0 keV.  
The parameter values that minimize $\chi^2$ are taken as the ``best fit".
The results are shown in Figures 6 and 7 and the
best-fit parameters are listed in Table~2. 
(The quoted errors in Table~2 
are 90\% confidence limits calculated for one variable.  Since
$\nh$ and $kT$ are not independent, the range of values and coupling is better
illustrated by the result of the joint analysis shown in Figure 7.)
In the three regions, the soft components have significantly
different $\nh$, (1.9--2.8)$\times 10^{22}$~cm$^{-2}$, but 
similar temperatures ($kT\simeq 0.3$~keV).
The different $\nh$ are consistent with
the variation of {\it total} column densities of
hydrogen nuclei along these directions,
which are listed in Table~2 for comparison. In Table~2,
$\nhone$\ is HI column density derived 
using $\it colden$ (a program
provided by {\em Chandra} X-ray center to compute total Galactic HI
column densities based on the compilation by Dickey \& Lockman [1990]),
while $\nhtwo$\ is
H$_2$ column density dervied from the CO data described in \S~3.2.
The agreement between the X-ray absorbing column densities and the
observed total column densities is very good. Therefore, we conclude that 
the three regions appear to be behind most of atomic
and molecular gases in the Sagittarius arm.

The parameters of the hard components in regions XN and XS
are poorly defined. Only the temperatures can be said to be
an order-of-magnitude
higher than the low-temperature component. Indeed, the hard X-rays 
probably originate from sources with different characteristics 
and the parameters in Table~2 
should not be taken at face value. But 
the parameters of the soft component in Table~2 are 
not sensitive to which model used for the hard energy part.
We will explore the spectral properties 
of the hard X-ray sources in \S~4.2.

\subsubsection{SIS Spectra}

SIS has higher energy resolution and can be used to determine
accurate parameters of emission lines.
However, in W51, the superposition of hard X-ray sources 
and the absence of a `clean' background
make the analysis difficult. 
We therefore limit the analysis to broadband model fitting.
Figure 8 shows the SIS spectra of regions XN and XS.
The areas used to derive the spectra are 
marked as solid circles in Figure 3a and do 
not include compact hard X-rays sources.
The background spectra were estimated from relatively emission-free regions in
the field. 
We fit a single-temperature, equilibrium thermal model to 
the spectrum of XN between
$E=0.7$ and 2.5 keV, because, at higher energies, there is a contamination 
by diffuse hard X-ray emitting gas.
We obtain
$\nh = 2.60 ^{+0.23}_{-0.24}\times 10^{22}$ cm$^{-2}$ and 
$kT = 0.25^{+0.05}_{-0.03}$ keV
($\chi^2=50/39$) and the result is shown in Figure~8. 
The temperature agrees 
with that derived from the GIS spectrum, 
while the absorbing column density is a little larger.
Similary, for region XS, we obtain
$\nh = 1.98\pm0.22\times 10^{22}$ cm$^{-2}$ and 
$kT = 0.34^{+0.09}_{-0.06}$ keV  
($\chi^2=30/23$) in agreement with the result from the GIS spectrum. 

\subsection{Hard X-ray Sources}
\subsubsection{HX1}

HX1 is an extended source associated with compact H II regions.
Figure~9 shows the spectrum of HX1. The background emission was
estimated from a source-free region in the southern part of the field.
The spectrum of HX1 has an indication of Si K line at $\sim 2.0$ keV (and perhaps 
S K line at $\sim2.4$ keV),  
which, together with the geometrical coincidence with the compact H~II regions, 
suggests that it is of thermal origin.
We fit a single-temperature, equilibrium thermal model to the energy
band $E=0.7$--8.0 keV. The result is shown in Figure~9 and the best-fit parameters are 
summarized in Table~3. In Table~3, $F_X$ and $L_X$ are 
the X-ray flux and X-ray luminosity between $E=0.7$ and 10.0 keV.

\subsubsection{HX2}

HX2 is unidentified. The absence of CO emission suggests that 
it is either a background source or possibly a source 
associated with the SNR W51C, e.g., a pulsar synchrotron nebula. 
Figure 10 shows the GIS and SIS spectra of HX2. 
The soft-energy spectra include a contribution from the emission of the SNR. 
The contamination might be smaller for the SIS spectrum because the instrument has 
a sharper point spread function and the spectrum was extracted from a 
region chosen to encompass only HX2.
(For the areas used to extract the spectra, see Figures 2c and 3b.) 
The GIS spectrum, on the other hand, has a higher signal-to-noise ratio in the 
hard energy band.  
We simultanesouly fit the GIS and SIS spectra with two components,
i.e., soft and hard components representing the emissions from the SNR and HX2, 
respectively. The soft component is assumed to be thermal and its 
parameters are fixed for those of region XS in Table 2. 
For the hard component, we 
consider both power-law and thermal emission models, but their 
absorbing column densities are assumed to be equal to 
that of XS ($1.86\times 10^{22}$~cm$^{-2}$), which is also 
equal to the total column density in this direction. This implicitly assumes that 
HX2 is either a background source or a source associated 
with the SNR. Even if not, the derived photon index and plasma temperature 
are not sensitive to absorbing column density.
Both models can fit the spectra with 
comparable statistical significance, and Table~3 summarizes the results. 
The power-law result is plotted in Figure 10. 

\subsubsection{HX3}

HX3 includes two point-like sources: 
HX3-east associated with a compact H II region and
HX3-west which is unidentified. 
Figure 11 shows the GIS and SIS spectra of HX3. 
The area used to extract the GIS spectrum includes 
both HX3-east and HX3-west, 
while the area for the SIS spectrum includes only HX3-east 
(see Figures 2c and 3b). 
We therefore fit the GIS and SIS spectra 
with three and two components, respectively.
Two components for the SIS spectrum are the soft and hard components 
representing the emissions from 
the SNR and HX3-east, respectively. For the GIS spectrum, we add 
one more component representing the hard X-ray emission from HX3-west.
Because of low S/N ratio, however, we want to fix the parameters of these 
components as much as possible. First in \S~4.1, we found that the SNR 
is probably at the far side of the Sagittarius arm. The column density toward 
the compact H~II region associated with HX3-east is also large (see \S~5.2).  
We do not have any information for HX3-west, but it is 
possibly a background source. 
We therefore fix the X-ray absorbing column density toward all the components for 
$\nh=3.0\times 10^{22}$~cm$^{-2}$, which 
is the total column density in this direction.
We adopt the best-fit temperature (0.29 keV) 
of region XN as the temperature of the SNR emission. The emission from HX3-east 
is assumed to be thermal because it is associated with a compact H~II region.
For HX3-west, we just consider power-law emission model.
The results are shown in Figure~11 and Table~3. The three-component fit to the GIS 
spectrum is, of course, not unique but results are consistent with our 
analysis of the region.
 
\section {DISCUSSION}
\subsection{W51C SNR}

According to our results,  
most parts of W51C emit X-rays from 
hot gas at a temperature of $\simeq$0.3~keV. 
(The temperature of the faint region along the
southeastern boundary of the GIS field is consistent with this too, i.e.,
$kT=0.18^{+0.19}_{-0.07}$ keV, although it has a large uncertainty.)
The possibility of the central X-ray emission being largely a synchrotron 
nebula, as considered by KKS, is ruled out. 
Furthermore, since the absorbing columns toward the X-ray emitting regions
are approximately equal to the total column densities of hydrogen nuclei in
those directions, the X-rays probably originate behind most of the atomic
and molecular gas in the Sagittarius arm. The {\em ASCA} results therefore
are compatible with W51C being a single, isothermal SNR at the far side of the
Sagittarius arm.

At a distance of 6 kpc, the 
linear extent of W51C is $88\times 66$~pc, which makes it 
one of the largest SNRs in the Galaxy (e.g., Strom 1996). 
We have derived the parameters of the central region; 
an elliptical ($23'\times 15'$) area surrounding regions XN and XS. 
The root mean, volume-averaged, square density 
$\bar n_e\equiv \left( \int n_e^2 dV/V \right)^{1/2}\simeq 0.58$~cm$^{-3}$, 
the X-ray emitting mass $M_X\simeq 290~M_\odot$, and 
the thermal energy $E_{\rm th}\simeq 0.39\times 10^{51}$~ergs. 
For comparison, KKS derived $\bar n_e\simeq 0.29$~cm$^{-3}$,
$M_X\simeq 1900~M_\odot$, and $E_{\rm th}\simeq 2.6\times 10^{51}$~ergs 
for the whole ($50'\times 38'$) SNR. 
If the X-ray emitting gas is clumpy with a volume filling factor of $f$, 
the true electron density $n_e$ would be greater by a factor of $f^{-1/2}$, whereas 
the X-ray emitting mass would be smaller by a factor of $f^{1/2}$. 
Thermal energy could be 
greater ($f^{-1/2}$) or smaller ($f^{1/2}$) depending on 
whether the interclump region is filled for the SNR 
to be isobaric or not.
From the {\em Einstein} image (Fig. 1), we estimate 
that the volumne filling factor of the X-ray emitting gas  
$f\simlt 0.4$ for the whole SNR, 
so that $n_e\simgt 0.46$~cm$^{-3}$, $M_X\simlt 1200~M_\odot$, and 
$E_{\rm th}\simgt 4.1\times 10^{51}$~ergs or 
$E_{\rm th}\simlt 1.6\times 10^{51}$~ergs. 
KKS estimated the age of the SNR at $\sim 3\times 10^4$~yrs.

As a single SNR, W51C has a composite morphology:  
a very thick ($\sim 13'$ or 23~pc) outer radio and X-ray 
shell and an X-ray--bright central region.
There are about 15 other composite SNRs, corresponding to $\sim 25$\% of 
X-ray--detected Galactic SNRs \citep{rho98}.
Several models have been suggested, 
which include 
the evaporation cloud model of \citet{whi91} and the heat conduction model of 
\citet{she99}.
The former assumes small dense clouds evaporating within 
the hot interior, while 
the latter considers only heat conduction within the hot interior. 
Observationally, many of these 
SNRs are known to be interacting with large molecular clouds 
(see Rho \& Petre 1998 and references therein). 
It has been shown that W51C is interacting with a molecular cloud 
along the western boundary of 
region XN \citep{koo97b}, and some, if not all, 
of the X-ray emitting gas in the central area
could have been originated from this molecular cloud. 

Since W51C is located in the Sagittarius arm and is interacting with a molecular cloud,
is likely to be a type II SNR. According to our results, however, 
most of the soft X-ray emission in the central region 
is thermal, certainly not consistent with a 
bright pulsar wind nebula. Among hard X-ray sources, 
HX2 and HX3-west are unidentified and could be candidates for a weak pulsar nebula. 
Their spectra can be fitted by a power-law model with $\Gamma=2.59^{+0.27}_{-0.29}$ 
and $1.7^{+1.0}_{-\infty}$ (or $\simlt 2.7$), 
respectively, although they can be fitted equally well by a thermal 
emission model. These photon indices fall into the range (1.6--2.5) of the indices of 
known pulsar nebulae \citep{sew88}. If we extraploate their spectra to 
low energies by the same power law, their X-ray 
luminosities in the {\em Einstein} band (0.2--4 keV) would be 
$\sim$0.4--2$\times 10^{34}$~ergs s$^{-1}$ assuming $d=6$~kpc. 
This is $\simlt 0.1$~\% of Crab nebula and is not unreasonable as a luminosity 
of a $\sim 3\times 10^4$~yrs old pulsar \citep{sew88}.
Recent {\em Chandra} discoveries also indicate that many young pulsars 
and pulsar wind nebulae do not have high luminosities \citep{olb01, hug01}.
High spatial resolution is needed 
in order to explore the nature of these sources and 
to look for a faint pulsar and its nebula.

\subsection{X-ray emission from W51 star-forming regions}

We have detected hard X-ray sources coincident with several 
compact H~II regions in W51.
The \easca hard image (Fig. 2c) implies that W51A might be the 
brightest hard X-ray source among the W51 star-forming regions, 
but the source is 
located at the edge of the GIS field and its properties could not be 
obtained. Among the W51B sources, 
two compact H II regions, G48.9$-$0.3 and G49.0$-$0.3, are coincident with 
the brightest hard X-ray source HX1. 
The brightest radio continuum source in W51B, G49.2$-$0.3,
is also detected as HX3-east, but is much fainter. 
These X-rays might be from early-type stars and young stellar objects 
in these compact H~II regions. 

Recent {\em Chandra} observation of the Orion Trapezium region showed that 
the X-rays are mostly from individual OB stars \citep{sch01}.
The three brightest stars, $\theta^1$ Ori A, C, \& E, of spectral type O7--B0.5 have
$L_X$=(2--19)$\times 10^{31}$ erg s$^{-1}$ and account for more than 80\% of the 
total luminosity of the Trapezium.
Stars of later spectral type are much fainter. 
The X-ray luminosity of HX1 
is $1.6\times 10^{34}$ erg s$^{-1}$, an order of 
magnitude greater than that of the Orion star-forming region \citep{fei00}.
According to \citet{han01}, who 
analyzed the 
{\em 2MASS} data of W51B to derive stellar populations of compact H~II regions, 
G48.9$-$0.3 and G49.0$-$0.3 together have 3 O-type stars (O6.5, O8, and O9), 8 B0 stars, 
and 13 B1 stars and these do not seem to be enough to account for the 
observed X-ray luminosity. It is possible that there are OB stars 
unidentified from the {\em 2MASS} data and/or some diffuse emission.
The {\em Einstein} observations of the Carina Nebula showed that 80\% of the 
X-ray emission came from a diffuse component \citep{sew82}.
Alternatively, the OB stars in these compact H~II regions 
could be brighter in X-rays than those in Orion. The X-ray luminosity 
of O-stars does seem to depend on their surroundings \citep{chl89}.
We need high-resolution X-ray observations to find out
the sources of X-rays in these compact regions.

One important result of the X-ray study is that we can determine 
the locations of compact H II regions within molecular clouds.
The X-ray absorbing column density toward HX1 
($\nh\simeq 1.5\times 10^{22}$~cm$^{-2}$) 
is not very much greater than the column density of {\it atomic}
hydrogen to the associated compact H II regions, i.e., 
$\nhone=$(1.23--1.50)$\times 10^{22}$~cm$^{-2}$ \citep{koo97}. 
Therefore, if we assume that most of the X-rays are from these H II regions, 
our result implies that the H II regions are in the front side of the molecular cloud.
This is consistent with the result of \citet{han01}. 
They found that the extinction toward exciting stars 
in G48.9$-$0.3 and G49.0$-$0.3 corresponds to $A_V\simeq 8$, 
which agrees very well with our X-ray absorbing column density if we consider 
$\nh/A_V\simeq 1.8\times 10^{21}$~cm$^{-2}$ mag$^{-1}$ \citep{sew00}.
For HX3-east, however, instead of deriving an X-ray absorbing column density from 
a model fitting, we adopted total, i.e., atomic and molecular, 
hydrogen column density because of the uncertainty due to other superposed sources.
According to \citet{han01}, G49.2$-$0.3 associated with HX3-east is 
located in the back side of the molecular cloud.

\section{CONCLUSIONS}

W51 is a complex region with 
a SNR and star-forming regions superposed within an area
of angular diameter $\sim 1^\circ$. 
Soft X-rays from the SNR were detected by {\em Einstein} and ROSAT, but,
because of the limited energy range of detector, it was not conclusive whether
the emission was thermal or non-thermal. The \easca study in this paper clearly
shows that most parts of the SNR emit thermal X-rays and the temperature of
the hot gas is about $0.3$~keV. It also shows that the X-rays originate behind
most of the atomic and molecular gas along those line of sights. These results
are compatible with X-rays being emitted from a single, isothermal SNR at the
far-side of the Sagittarius arm, which makes the W51C SNR one of the largest
SNRs in the Galaxy.

The \easca observations reveal hard X-rays from 
the W51B star-forming regions. 
Two regions (G48.9$-$0.3 and G49.0$-$0.3) appear 
bright in hard X-rays and it is likely that they are located  
in the front side of the molecular cloud. 
W51 is a rare example of a SNR superposed on a star-forming region.
Since the progenitors of Type II SN are massive stars, this 
situation should be more common, but it is not.
Apparently W51 is one of those few regions in the Galaxy 
where we can study the formation of massive stars and the consequence of 
their violent explosions simultaneously. High-resolution X-ray observations 
are needed to fully identify the origins of the central emission.

\acknowledgements
We are grateful to Miller Goss, Ravi Subrahmanyan, and John Carpenter 
for providing their H~I and CO data for comparison with the \easca data.
We wish to thank Nancy Brickhouse for providing a list of strong lines
in X-rays. We also wish to thank the anonymous referee for helpful comments.
BCK wish to thank Chul-Sung Choi for helpful discussions. 
JJL would like to thank the Ministry of Education for its financial support. 
This work was supported by the Korea Research Foundation Grant 
(KRF-2000-015-DP0446). 
{}

\clearpage

\figcaption[]{Position of {\em ASCA} SIS (squares) and GIS (circle) 
fields marked on the ROSAT image of Koo et al. (1995). 
Contour levels are 
linearly spaced from 10\% to 80\% of the peak brightness 
with steps of 10\%. Peak 
brightness excluding the strong point source in the field 
is $1.2\times 10^{-4}$ counts s$^{-1}$ pixel$^{-1}$.} 

\figcaption[]{GIS2+3 images in the 
(a) 0.7--10.0 keV, (b) 0.7--2.5 keV,  and (c) 2.5--6.0 keV energy bands. 
Background subtraction, vignetting correction, and exposure correction 
were made. Contour levels are 
linearly spaced from 20\% to 90\% of the peak brightness 
with steps of 10\% in each map. Peak 
values are 1.85, 1.47, and 0.90$\times 10^{-4}$ 
counts s$^{-1}$ pixel$^{-1}$, respectively. 
The solid circles mark 
the areas used for spectral analysis.
(d) X-ray color map generated from three energy bands;
0.7--1.5 keV (red), 1.5--2.5 keV (green), and 2.5--6.0 keV (blue).}

\figcaption[]{SIS0+1 images in the (a) 0.7--2.5 keV and (b) 2.5--6.0 keV
energy bands. 
Vignetting and exposure corrections were made. 
Contour levels are
linearly spaced from 30\% to 90\% of the peak brightness 
with steps of 10\% in each map. Peak
values are 0.17 and 0.24$\times 10^{-4}$ 
counts s$^{-1}$ pixel$^{-1}$, respectively. The crosses in (a) mark the peak positions of 
XN and XS in Figure 2b, while those in (b) mark the peak positions of 
HX2 and HX3-east in Figure 2c.
The solid circle and ellipses mark 
the areas used for spectral analysis.}

\figcaption[]{330~MHz radio contour map of the W51 complex overlaid on 
the (a) ROSAT and (b) \easca GIS hard (2.5--6.0 keV) 
X-ray images.  
The ROSAT X-ray map is used instead of the \easca GIS 
soft X-ray image because it covers a larger area. 
The 330 MHz map is from Shurirahmanian \& Goss (1995). 
In radio, W51 is composed of two complex H~II regions, W51A and W51B, and the 
shell-type SNR W51C. Their approximate positions are marked in (a).  
Compact radio sources labeled in (b) are the W51B H~II regions which 
are associated with hard X-ray sources. }

\figcaption []{Same as Figure 4, but with 
$^{12}$CO integrated intensity contour map. 
The CO map is obtained integrating over the velocity range of 0 to +75~\kms\ 
and shows all the molecular gas toward this direction.
The contour levels (in K \kms) 
are 20, 40, 60, 80, 100, 150, and 200. 
The column density of hydrogen nuclei corresponding to 1 K~\kms\ is 
about $4.6\times 10^{20}$~cm$^{-2}$. The CO maps are generated from 
the data of Carpenter \& Sanders (1998).}

\figcaption[] {GIS spectra of regions XN, XS, and XW. The solid lines show the 
best-fit equilibrium thermal emission models (MEKAL).
For regions XN and XS, a two-temperature model was fit  
for $E=0.7$--6.0 keV. For region XW, a single-temperature 
model was fit for $E=0.7$--2.0 keV. 
Dashed lines show the contributions from individual components.}

\figcaption[] {90\% confidence contours for the soft-component parameters of 
regions XN, XS, and XW.}

\figcaption[] {Same as Figure 6, but for the 
SIS spectra of regions XN and XS.}

\figcaption[] {GIS spectrum of HX1. The solid line shows the
best-fit equilibrium, thermal emission model.}

\figcaption[] {GIS and SIS spectra of HX2. The solid lines show the 
best-fit two component model with soft and hard components representing the emissions  
from the SNR and HX2, respectively. Dashed lines show the contributions 
from individual components.}

\figcaption[] {GIS and SIS spectra of HX3. The SIS spectrum is fitted with 
two components with soft and hard components 
representing the emission from the SNR and HX3-east, respectively. 
For the GIS spectrum, we added one more component (dotted line) representing HX3-west.}

\clearpage

\begin{deluxetable}{lcclccc}
\tabletypesize{\scriptsize}
\tablecaption{Properties of Hard X-ray Sources \label{tbl-1}}
\tablewidth{0pt}
\tablecolumns{7}
\tablehead{
\colhead {} &  \colhead {} & \colhead {} 
&\multicolumn{4}{c}{Radio Counterpart}\\
\cline{4-7}\\
\colhead {} & \colhead {Peak Position} & \colhead {Count Rates$^{\rm a}$} & \colhead {} 
& \colhead {} & \colhead {21-cm Flux Density} & \colhead {} \\
\colhead {Source} & \colhead { ({$\alpha_{\rm 2000}$}, {$\delta_{\rm 2000}$}) } 
& \colhead {($\times 10^{-2}$ counts s$^{-1}$)} & \colhead {Source}  
& \colhead { ({$\alpha_{\rm 2000}$}, {$\delta_{\rm 2000}$}) } 
& \colhead {(Jy)} & \colhead {Reference} }

\startdata
HX1 & (19 22 21, 14 05 30) &4.3 & G48.9-0.3 & (19 22 14.2, 14 02 57) & 4.02 & 1 \\
& & & G49.0-0.3 & (19 22 25.7, 14 06 18) & 1.23 & 1 \\
HX2 & (19 23 17, 14 02 50) & 1.1 & unidentified & (19 23 18.3, 14 02 31) & 0.093 &2 \\
HX3-east & (19 23 02, 14 16 30)  &0.8 &G49.2-0.3& (19 23 01.3, 14 16 50) & 7.42 & 1 \\
HX3-west & (19 22 47, 14 16 20) & 1.0 & unidentified & (19 22 45.5, 14 16 06) & 0.024 &2 \\
\enddata

\tablenotetext{a} {Between $E=2.5$ and 6.0 keV.}
\tablerefs {(1) Koo 1997; (2) Condon et al. 1998.}

\end{deluxetable}
\clearpage

\begin{deluxetable}{ccccccccc}
\tablecaption{Spectral Analysis of the Regions Bright in Soft X-rays \label{tbl-2}}
\tabletypesize{\scriptsize}
\tablewidth{0pt}
\tablecolumns{9}
\tablehead{
\colhead {} & \multicolumn{2}{c}{Soft Component} & 
\multicolumn{2}{c}{Hard Component} & \colhead {} & 
\multicolumn{3}{c}{Atomic and Molecular Columns} \\
\cline{2-3} \cline{4-5} \cline{7-9}\\
\colhead {} & \colhead {$\nh$} & \colhead {$kT$} & \colhead {$\nh$} &\colhead {$kT$}
&\colhead {} &\colhead{$\nhone$} & \colhead{$2\nhtwo$} & \colhead{$\nhone+2\nhtwo$}\\
\colhead {Region} & \colhead { (10$^{22}$ cm$^{-2}$)} & \colhead { (keV)}
&\colhead {(10$^{22}$ cm$^{-2}$)} &\colhead {(keV)} &\colhead {$\chi^2$}  & \colhead { (10$^{22}$ cm$^{-2}$)}  & \colhead { (10$^{22}$ cm$^{-2}$)}  & \colhead { (10$^{22}$ cm$^{-2}$)} 
}

\startdata
XN & $2.25^{+0.20}_{-0.17}$ & $0.29^{+0.06}_{-0.05}$ & $1.8^{+1.9}_{-1.2}$ & 
$2.5^{+2.0}_{-0.8}$ & 111/102 & 1.5 & 0.9 & 2.4 \\
XS & $1.86^{+0.38}_{-0.22}$ & $0.34^{+0.05}_{-0.06}$ & $\simlt 4.4$ 
& $\simgt 2.4$ & 78/92 &1.4 & 0.4 & 1.8 \\
XW & $2.8^{+0.9}_{-0.6}$ & $0.29^{+0.14}_{-0.10}$ & ... & ... & 29/23 & 1.6 & 1.1 & 2.7 \\

\enddata

\end{deluxetable}
\clearpage

\begin{deluxetable}{llllccl}
\tabletypesize{\scriptsize}
\tablecaption{Spectral Analysis of Hard X-ray Sources \label{tbl-3}}
\tablewidth{0pt}
\tablecolumns{7}
\tablehead{
\colhead {} & \colhead {Emission} & \colhead {$\nh$} & \colhead {$kT$} & 
\colhead {} & \colhead{${F_X}^{\rm a}$} & \colhead{${L_X}^{\rm a,b}$}\\
\colhead {Source} & \colhead {Model} & \colhead { (10$^{22}$ cm$^{-2}$)} 
& \colhead { (keV)} & \colhead {$\Gamma$} & \colhead 
{($10^{-12}$ ergs cm$^{-2}$ s$^{-1}$)} &\colhead {($10^{34}$ ergs s$^{-1}$)}
}

\startdata
HX1 & thermal & $1.45^{+0.31}_{-0.30}$ & $3.3^{+1.1}_{-0.6}$ & ... & 
5.5 & 1.6 \\
HX2 & power-law & 1.86$^{\rm c}$ & ... & $2.59^{+0.27}_{-0.29}$ & 
2.1 & 0.89 \\
& thermal & 1.86$^{\rm c}$ & $2.6^{+0.6}_{-0.4}$ & ... &
1.6 & 0.69 \\
HX3-east & thermal & 3.0$^{\rm c}$ & $\simlt 3.0^{\rm d}$ & ... &
1.5 & 0.45 \\
HX3-west & power-law & 3.0$^{\rm c}$ & ... & $\simlt 2.7^{\rm d}$ &
1.2 & 0.52 \\
\enddata

\tablenotetext{a} {Between $E=0.7$ and 10.0 keV.}
\tablenotetext{b} {HX1 and HX3-east are associated with compact H~II 
regions in W51B and their distances are assumed to be 5~kpc.
HX2 and HX3-west are unidentified sources and their distances are 
assumed to be 6~kpc.}
\tablenotetext{c} {These are fixed.}
\tablenotetext{d} {These parameters do not have a lower bound for the 90\% confidence interval.
${F_X}$ (and ${L_X}$) of HX3-east and HX3-west are based on the 
best-fit parameter values $kT=1.8$ keV and $\Gamma=1.7$, respectively.}

\end{deluxetable}

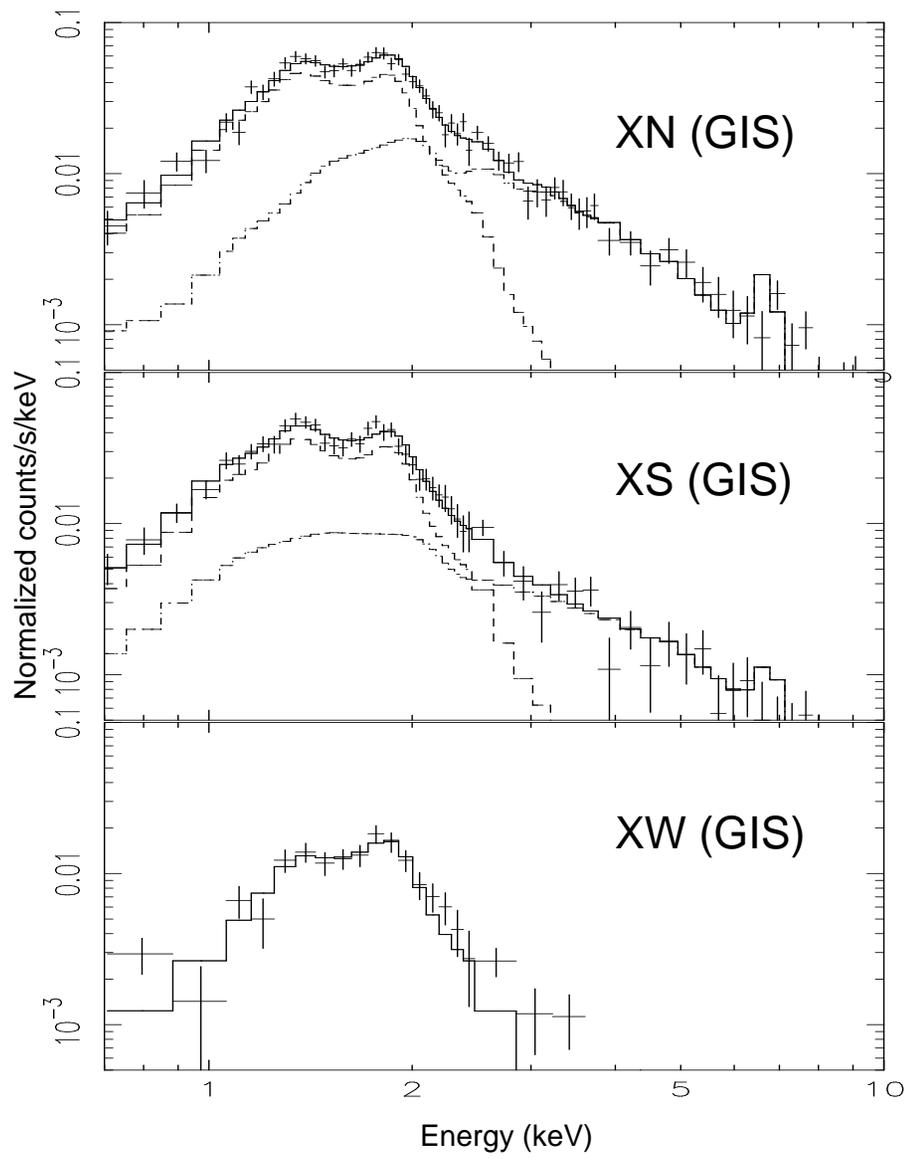
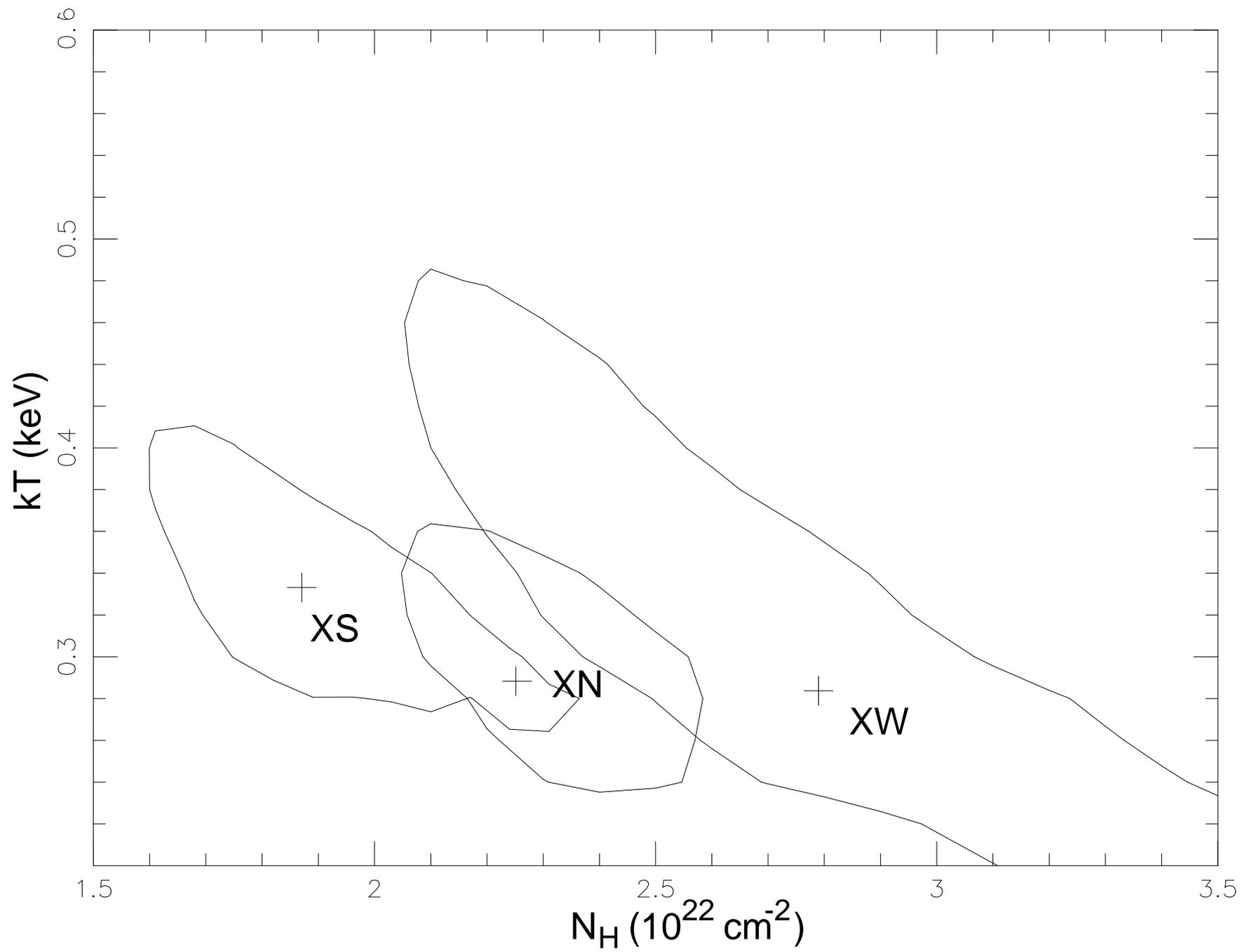
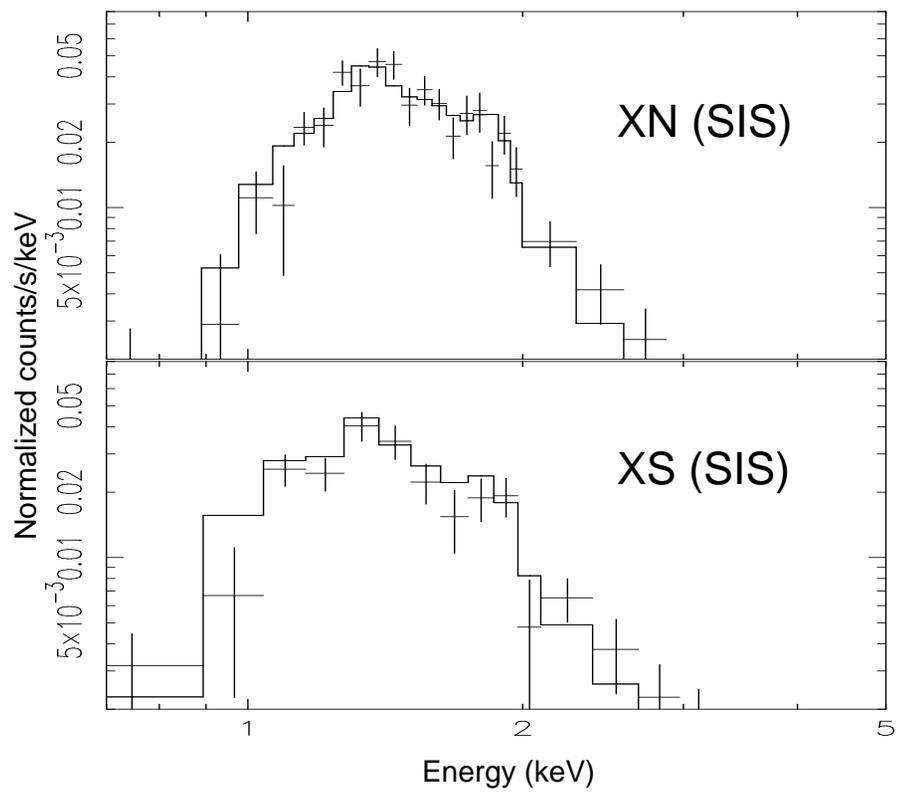
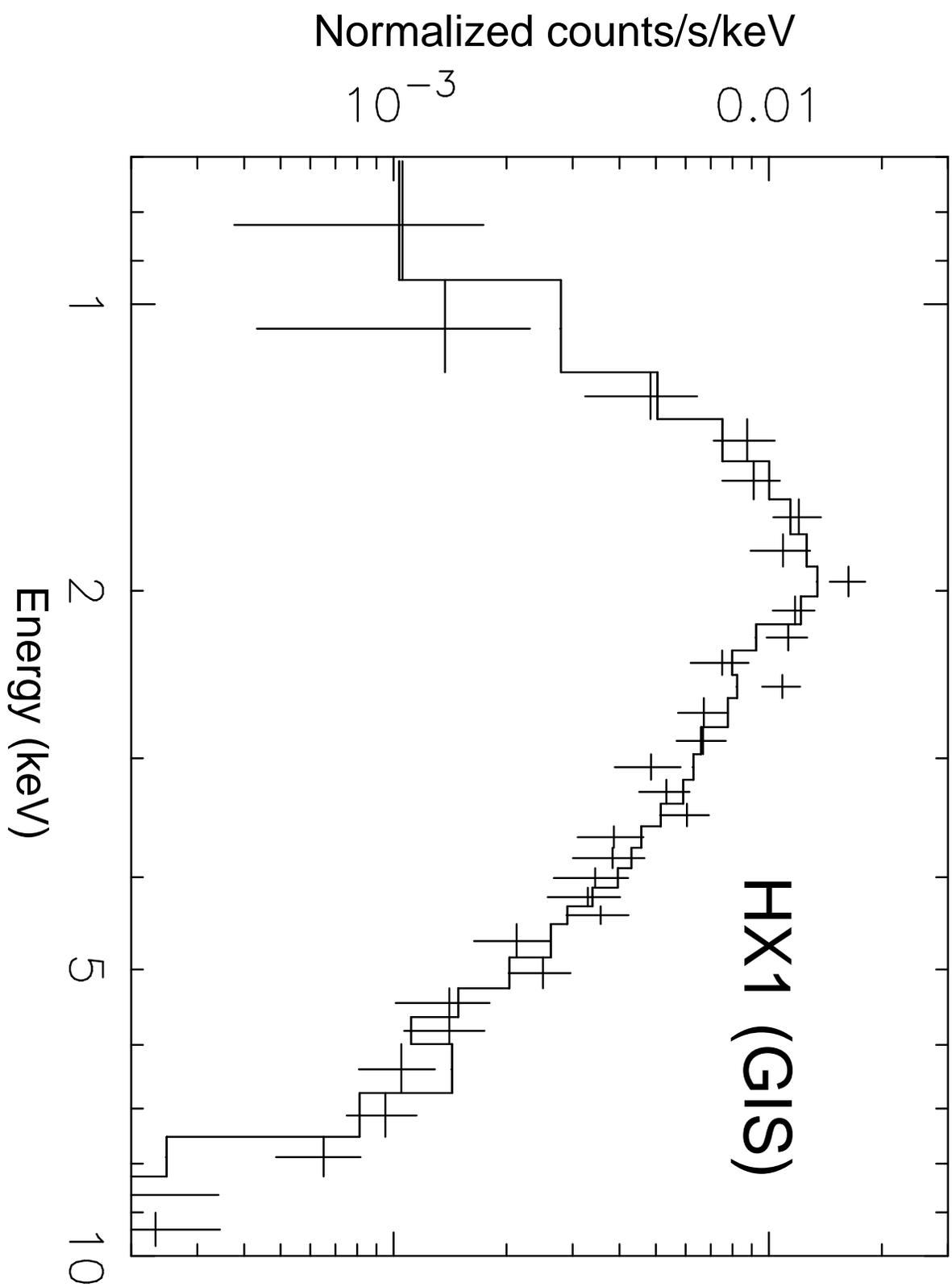
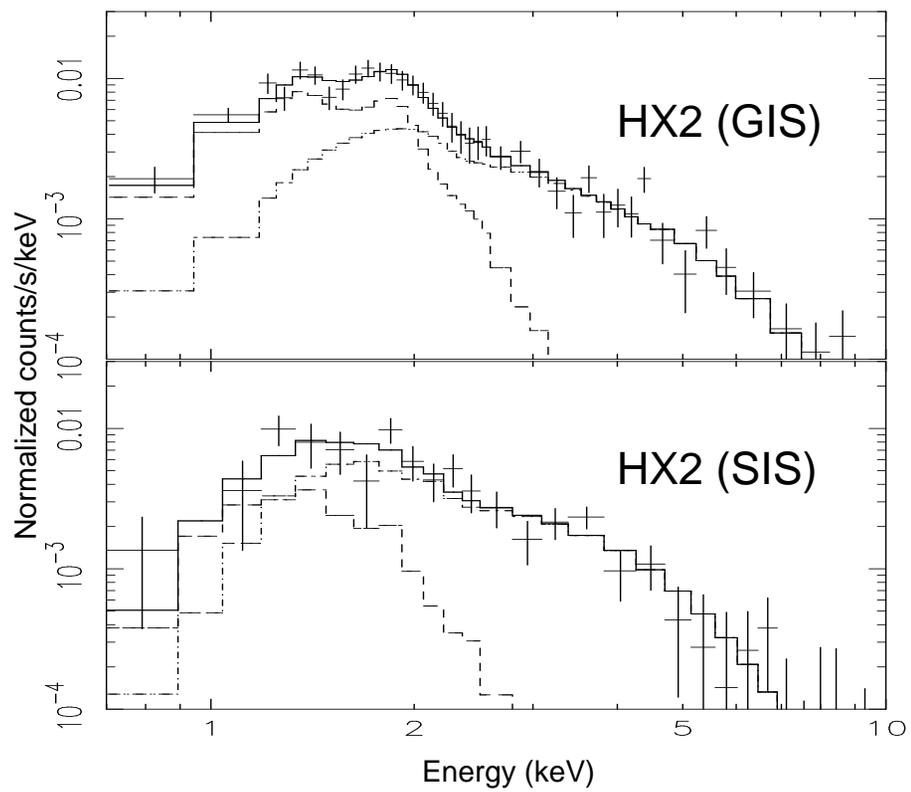
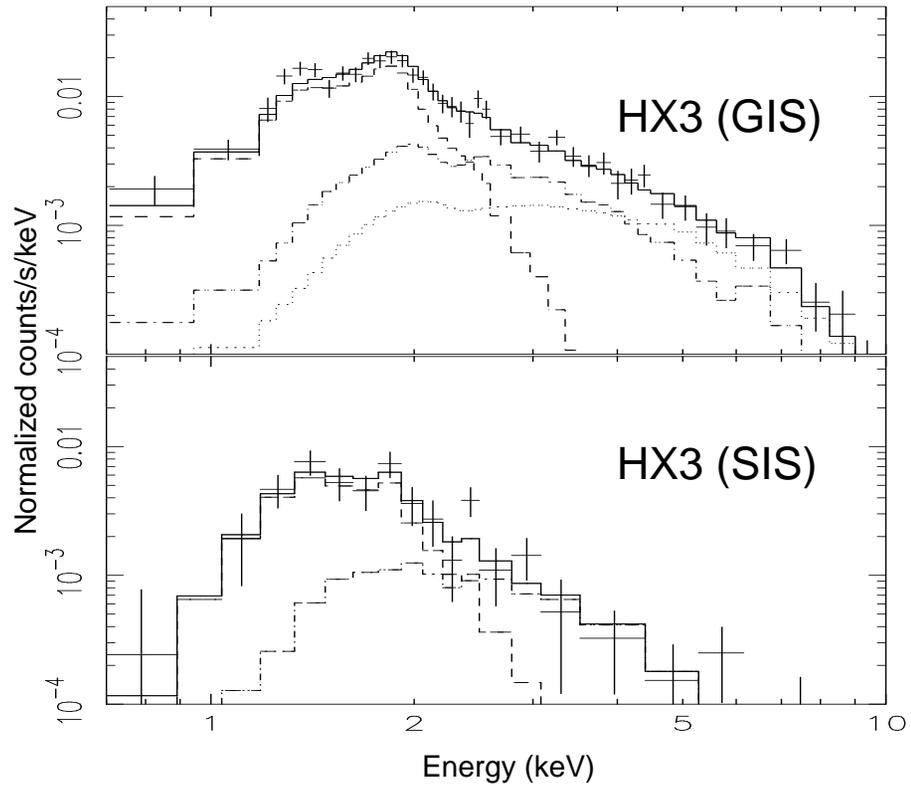
\end{document}